\magnification=1200 

\hsize= 6.35truein
\vsize=8.5truein

\def\nl{\par\noindent}

\pretolerance=10000

\nopagenumbers
\headline={\ifnum \pageno=1 \hfil \else\hss\hskip44.8pt\tenrm- \folio\
-\hss\fi {\tt
CBPF-NF-045/96}}

\font\ninerm=cmr9

\def\({\c c}
\def\|{\'\i}

\baselineskip=12pt 
\nopagenumbers

\def\nl{\par\noindent}

\pretolerance=10000

\centerline {{\bf Light-Front Quantization of Field Theory}
\footnote{$^{\dag}$}{\ninerm {\it Invited talk given at the  
{\sl Theoretical Physics Symposium } 
convened to celebrate the seventieth birthday of {\sl Paulo Leal Ferreira} 
at {\sl Instituto de F\|sica Te\'orica, UNESP}, 
S\~ao Paulo, August 1995. {\sl Topics in Theoretical Physics}, {\rm
pgs. 206-217, eds., V.C. Aguilera-Navarro et. al., IFT-S\~ao Paulo,
SP, 1995} }}} 

\medskip

\centerline {Prem P. Srivastava\footnote{$^{\star}$}
 {\ninerm  E-Mail: prem@vax.fis.uerj.br or @lafexsu1.lafex.cbpf.br}}
\medskip
\centerline {\it  Instituto de F\|sica, Universidade do Estado de  
Rio de Janeiro, }
\centerline {\it Rua S\~ao Francisco Xavier 524,  
Rio de Janeiro, RJ, 20550-013, Brasil. }
\centerline {\it and}
\centerline {\it Centro Brasileiro de Pesquisas F\|sicas - CBPF/LAFEX}
\centerline {\it Rua Xavier Sigaud, 150, Rio de Janeiro, RJ, 22290, Brasil}

\vskip 1cm
\centerline {\bf Abstract}
\medskip

{\leftskip 32pt\rightskip 32pt {\ninerm { Some basic topics in 
Light-Front (LF) quantized field theory are reviewed. Poincar\`e algebra and 
the {\it LF Spin operator } are discussed. The local scalar field 
theory of the conventional framework is shown to correspond to a 
{\it non-local  Hamiltonian} theory on the LF in view of the 
{\it constraint 
equations} on the phase space, which relate the bosonic condensates to  the
non-zero modes. This new ingredient is useful to describe the  
{\it spontaneous symmetry breaking} on the LF. The instability of the 
symmetric phase in two dimensional scalar theory when the coupling constant 
grows is shown in the LF theory renormalized to one loop order. 
{\it Chern-Simons gauge theory}, 
regarded to describe excitations with fractional statistics,  is quantized  
in the light-cone gauge and a simple LF Hamiltonian obtained which may 
allow us to construct renormalized theory of {\it anyons}.} }\par}


\bigskip
\nl {IF-UERJ 033/96}

\eject
\baselineskip=18pt
\nl {\bf 1 \quad Introduction}
\medskip

Dirac$^{1}$ in 1949 pointed out the advantage of studying the relativistic 
quantum dynamics of  physical system on the hyperplanes of the 
LF: $x^{0}+x^{3}=const.$, {\it front form}.   
Seven out of the ten   
Poincar\'e generators are here {\it kinematical } while in the 
conventional formulation on the hyperplanes $x^{0}=const$., 
{\it instant form}, only six have this property. 
LF field theory was rediscovered in 1966 by Weinberg$^{2}$ in his 
Feynman rules adapted for infinite momentum frame. It was
demonstrated$^{3}$  latter  that the rules correspond to 
the quantization on the LF. 

The LF vacuum is  simpler than the 
conventional theory vacuum and in many cases the interacting theory vacuum 
may coincide with the perturbation theory one. 
This results from the fact that  momentum 
four-vector  is now given by $(k^{-},k^{+},k^{\perp})$ where 
$k^{\pm}=(k^{0}{\pm}k^{3})/{\sqrt 2}$.  Here 
$k^{-}$ is the LF energy while $k^{\perp}$ and  $k^{+}$  indicate 
the transverse and the longitudinal components of the  momentum. 
For a massive particle on the mass shell   $k^{\pm}$ are positive definite 
and the conservation of the total 
longitudinal momentum does not permit the excitation of these quanta 
by the LF vacuum. 
The recent revival$^{4,5,6}$ of the interest in LF quantization 
owes to the difficulties encountered in the computation of 
nonperturbative effects, say, in the {\it instant form} QCD.  
In the  conventional framework QCD vacuum state is quite complex 
due to the {\sl infrared slavery} and  it contains  
also gluonic and fermionic condensates.  
There seems to exist  contradiction between 
the Standard Quark Model and the 
QCD containing quark and and gluon fields. 
Also in the Lattice gauge theory there is the  well known 
difficulty in handling light fermions. 
LF quantization  may throw some 
light to clarify this and other  issues. 
In the context of the String theories it has been 
used, for example,  in the case of the heterotic strings$^{7}$.

It is convenient to use LF coordinates corresponding to 
$\,(x^{0},x^{1},x^{2},x^{3})$ which are defined by 
$(x^{+},x^{-},x^{\perp})$ where $x^{\pm}=(x^{0}{\pm} x^{3})
/{\sqrt 2}=x_{\mp}$ and   $ x^{\perp}\equiv {\bar x} : 
(x^{1}=-x_{1},\,x^{2}=-x_{2})$. We will take $x^{+}\equiv \tau$ as the 
LF {\sl time} coordinate and  $x^{-}\equiv x$ as the longitudinal 
spatial coordinate\footnote{$^{\ddag}$}{\ninerm   Theory quantized,
say, at equal $x^{+}$  seems already to carry information on equal $x^{-}$ 
commutators as well and the role of $x^{+}$ and 
$x^{-}$ may be interchanged (See Appendix).}. 
The LF components of any four-vector or 
any tensor are similarly defined. The metric tensor for the indices 
$\mu=(+,-,1,2)$ is  $g^{++}=g^{--}=g^{12}=g^{21}=0; \,
g^{+-}=g^{-+}=-g^{11}=-g^{22}=1$. 
The transformation from the conventional to LF 
coordinates is seen {\it not} to be a Lorentz transformation. 

Any two non-coincident points on the hyperplane $x^{0}=const.$  
 have a spacelike separation: $(x-y)^{2}\vert_{x^{0}=y^{0}}=- 
(\vec x-\vec y)^{2}<0\,$ and it becomes lightlike when the points 
coincide. The points on the LF hyperplane  
$x^{+}=const.$  also have a spacelike separation: 
$(x-y)^{2}\vert_{x^{+}=y^{+}}=- 
(x^{\perp}-y^{\perp})^{2}<0\,$ which reduces to lightlike  when 
$x^{\perp}=y^{\perp}$, {\it but} with the important difference that now 
the points need not be necessarily coincident since 
$(x^{-}-y^{-})$ may take arbitrary value. 
Admitting also the validity of the {\it microscopic causality} principle 
it can be shown that  the appearence of nonlocality  in the LF field theory 
along the longitudinal direction $x^{-}$ is not necessarily 
unexpected. Consider, for example, the commutator 
$[A(x^{+},x^{-},{x^{\perp}}),B(0,0,0^{\perp})]_{x^{+}=0}$
of two scalar observables $A$ and $B$. The {\it microcausality} would
require it to vanish for  $ x^{\perp}\ne 0$ when $x^{2}\vert_{x^{+}=0}$ 
is spacelike. 
Consequently 
it is proportional to  $\,\delta^{2}(\bar x)\, $ and its derivatives 
which implies locality in $x^{\perp}$; however, no restriction on the 
$x^{-}$ dependence follows. Similar arguments in 
the equal-time case lead to the locality 
in all the three space coordinates. We note also that in 
view of the {\it microcausality} both $[A(x),B(0]_{x^{+}=0}$ and 
$[A(x),B(0)]_{x^{0}=0}$ may be nonvanishing 
only on the light cone $\,x^{2}=0\,$.

It is interesting to consider the Lehman spectral representation$^{8}$ 
for the scalar field 

$${\langle\vert [\phi(x),\phi(0)]\vert \rangle}_{0}= \int_{0}^{\infty}\,
d\sigma^2\,\rho(\sigma^2)\,\triangle(x;\sigma^2),\,\;
\triangle(x;\sigma^2)= \int_{-\infty}^{\infty}\,
 {d^{4}k\over {(2\pi)^3}}\epsilon(k^{0})
\delta(k^2-\sigma^2)\,e^{-ik.x} $$

\nl Here the spectral function  $\rho(\sigma^2)$ is Lorentz invariant 
and positive definite and $\triangle(x;\sigma^2)$ is the vacuum expectation 
value (v.e.v.) of the commutator of the free field and 
$\epsilon(y)=-\epsilon(-y)=\theta(y)-\theta(-y)
=1 $ for $y>0$. For the field theory with a local Lagrangian 
it can be shown in the equal-time framework that 
$\int_{0}^{\infty}d\sigma^2\,\rho(\sigma^2)=1\,$. On the LF, 
$d^{4}k=d^{2}{\bar k}dk^{+}dk^{-},\,k^2=2k^{+}k^{-}-{k^{\perp}}^2, 
\, k.x=k^{+}x^{-}+k^{-}x^{+}-{ k^{\perp}}.{x^{\perp}},\,$ and  
$(2\vert k^{+}\vert)\delta(k^2-\sigma^2)=
\delta(k^{-}-[{\bar k}^2+\sigma^2]/(2k^{+}))$. Hence we show 
that $\triangle(x^{+},x^{-},{\bar x};\sigma^2)\vert_{x^{+}=0}
=-{i\over 4}\delta^{2}(\bar x)\epsilon(x^{-})\,$ and it follows that on the
LF   
$ [\phi(x^{+},x^{-},{\bar x}),\phi(0)]\vert_{x^{+}=0}=
-{i\over 4}\delta^{2}(\bar x)\epsilon(x^{-})$ where v.e.v. of the 
expression is understood. In contrast to the equal-time case 
the equal-$\tau$  commutator is not vanishing and it has a nonlocal 
dependence on $x^{-}$. The same result will be shown to follow also in the 
canonical quantization on the LF when we use  the Dirac 
procedure$^{9}$   in order 
to construct the Hamiltonian framework.  
We remind that any field theory written in terms of the LF coordinates 
describes necessarily a 
constrained dynamical system with a singular Lagrangian. 

We remark that in the LF quantization we (time) order with 
respect to  $x^{+}$ rather than  $x^{0}$. The {\it microcausality}, 
however, ensures that the retarded commutators 
$[A(x),B(0)]\theta(x^{0})$ and  $[A(x),B(0)]\theta(x^{+})$ 
do not  lead to disagreement in the two formulations. 
In fact in the regions  
$x^{0}>0, x^{+}<0$ and $x^{0}<0, x^{+}>0$, where the commutators  
appear to give different values the $x^{2}$  is spacelike 
and consequently both of them vanish. Such (retarded) 
commutators in fact appear in the S-matrix elements when we use the 
 Lehmann, Symanzik and 
Zimmerman (LSZ)$^{10}$ reduction formulae.

\baselineskip=18pt

\bigskip
\nl {\bf 2. \quad Poincare Generators on the LF}
\medskip 

The Poincar\'e generators in coordinate system 
$\,(x^{0},x^{1},x^{2},x^{3})$,  satisfy 
$[M_{\mu\nu},P_{\sigma}]=-i(P_{\mu}g_{\nu\sigma}-P_{\nu}g_{\mu\sigma})$ and 
$[M_{\mu\nu},M_{\rho\sigma}]=i(M_{\mu\rho}g_{\nu\sigma}+
M_{\nu\sigma}g_{\mu\rho}-M_{\nu\rho}g_{\mu\sigma}-M_{\mu\sigma}g_{\nu\rho})
\,$  where the metric is 
 $g_{\mu\nu}=diag\,(1,-1,-1,-1)$, $\mu=(0,1,2,3)$ and we take 
 $\epsilon_{0123}=\epsilon_{-+12}=1$. If we define 
 $J_{i}=-(1/2)\epsilon_{ikl}M^{kl}$
and  $K_{i}=M_{0i}$, where   $i,j,k,l=1,2,3$, we find 
$[J_{i},F_{j}]=i\epsilon_{ijk}F_{k}\,$ 
for $\,F_{l}=J_{l},P_{l}$ or  $ K_{l}$ while 
$[K_{i},K_{j}]=-i \epsilon_{ijk}J_{k}, \, [K_{i},P_{l}]=-iP_{0}g_{il},
\, [K_{i},P_{0}]=iP_{i},$ and  $[J_{i},P_{0}]=0$.

 The LF  generators are 
$\,P_{+}, P_{-},P_{1},P_{2}\,$, $M_{12}=-J_{3},\,M_{+-}=-K_{3},\,M_{1-}=
-(K_{1}+J_{2})/{\sqrt 2}\,
\equiv {-B_{1}}, \; M_{2-}=-(K_{2}-J_{1})/{\sqrt 2}
\equiv {-B_{2}}, \;
M_{1+}=-(K_{1}-J_{2})/{\sqrt 2}\equiv -S_{1},\,$ and 
$M_{2+}=-(K_{2}+J_{1})/{\sqrt 2}\equiv -S_{2}\,$. We find 
$[B_{1},B_{2}]=0,  [B_{a},J_{3}]=-i\epsilon_{ab} B_{b}, 
[B_{a},K_{3}]=iB_{a}, [J_{3},K_{3}]=0, 
[S_{1},S_{2}]=0, [S_{a},J_{3}]=-i\epsilon_{ab} S_{b}, 
[S_{a},K_{3}]=-iS_{a}$ where $a,b=1,2$ and $\epsilon_{12}=-\epsilon_{21}=1$. 
Also $[B_{1},P_{1}]=[B_{2},P_{2}]=i P^{+}, [B_{1},P_{2}]=[B_{2},P_{1}]=
0,  [B_{a},P^{-}]=iP_{a},  [B_{a},P^{+}]=0, 
[S_{1},P_{1}]=[S_{2},P_{2}]=i P^{-}, [S_{1},P_{2}]=[S_{2},P_{1}]=
0,  [S_{a},P^{+}]=iP_{a},  [S_{a},P^{-}]=0,  
[B_{1},S_{2}]=- [B_{2},S_{2}]=-iJ_{3}, 
[B_{1},S_{1}]=[B_{2},S_{2}]=-iK_{3} $. 
For $ P_{\mu}=i\partial_{\mu}$,  and 
$M_{\mu\nu}\to L_{\mu\nu}= i(x_{\mu}\partial_{\nu}-x_{\nu}\partial_{\mu})$ 
we find 
$B_{a}=(x^{+}P^{a}-x^{a}P^{+}), S_{a}=(x^{-}P^{a}-x^{a}P^{-}), 
 K_{3}=(x^{-}P^{+}-x^{+}P^{-})$  and 
$  \quad J_{3}=(x^{1}P^{2}-x^{2}P^{1})$. 
Under the conventional {\it parity} operation  ${\cal P}$: 
($\;x^{\pm}\leftrightarrow x^{\mp}, x^{1,2}\to 
-x^{1,2}$) and $(  p^{\pm}\leftrightarrow p^{\mp}, p^{1,2}\to 
-p^{1,2}),$ we find  $\vec J\to \vec J,\, \vec K \to -\vec K $, $B_{a}\to -S_{a}$  
etc..
The  six generators 
$\,P_{l}, \, M_{kl}\,$  leave  $x^{0}=0$ hyperplane 
invariant and are  called$^{1}$  
{\it kinematical}  while the remaining $P_{0},\,M_{0k}$ 
the  {\it dynamical } ones. On the LF 
there are {\it seven} kinematical generators :  $P^{+},P^{1},
P^{2}, B_{1}, B_{2}, J_{3}$ and  $K_{3} $ which leave the LF hyperplane, 
$x^{0}+x^{3}=0$,  invariant and the three {\it dynamical} 
ones $S_{1},S_{2}$ and  $P^{-}$ 
form a mutually commuting set. We note that each of the set 
 $\{B_{1},B_{2},J_{3}\}$ and   $\{S_{1},S_{2},J_{3}\}$ generates 
an  $E_{2}\simeq SO(2)\otimes T_{2} $ algebra; this will be shown below to be 
relevant for 
defining the {\it spin} for massless particle. Including $K_{3}$ in each set 
we find two subalgebras each with four elements. Some useful identities are 
$e^{i\omega K_{3}}\,P^{\pm}\,e^{-i\omega K_{3}}= e^{\pm \omega}\,P^{\pm},
\, e^{i\omega K_{3}}\,P^{\perp}\,e^{-i\omega K_{3}}= P^{\perp}, 
 e^{i\bar v.\bar B}\,P^{-}\,e^{-i\bar v.\bar B}= 
 P^{-}+\bar v.\bar P + {1\over 2}{\bar v}^{2}P^{+},  
 e^{i\bar v.\bar B}\,P^{+}\,e^{-i\bar v.\bar B}= P^{+}, 
 e^{i\bar v.\bar B}\,P^{\perp}\,e^{-i\bar v.\bar B}= 
 P^{\perp}+v^{\perp} P^{+}, 
 e^{i\bar u.\bar S}\,P^{+}\,e^{-i\bar u.\bar S}= 
 P^{+}+\bar u.\bar P + {1\over 2}{\bar u}^{2}P^{-}, 
  e^{i\bar u.\bar S}\,P^{-}\,e^{-i\bar u.\bar S}= P^{-}, 
 e^{i\bar u.\bar S}\,P^{\perp}\,e^{-i\bar u.\bar S}= 
 P^{\perp}+u^{\perp} P^{-}$ 
  where  $P^{\perp}\equiv\bar P=(P^{1},P^{2}), \, v^{\perp}\equiv {\bar v}
= (v_{1},v_{2})\, $ 
and  $(v^{\perp}. P^{\perp})\equiv(\bar v.\bar P)=v_{1}P^{1}+v_{2}P^{2}$ etc. Analogous expressions with 
$P^{\mu}$  replaced by $X^{\mu}$ can be obtained if we use 
$\,[P^{\mu},X_{\nu}]\equiv [i\partial^{\mu},x_{\nu}]= i\delta^{\mu}_{\nu}\,$. 

\bigskip

\bigskip
\nl {\bf 3. \quad  LF Spin Operator. Hadrons in LF Fock Basis }
\medskip 

\bigskip

The Casimir generators of the Poincar\'e group are : $P^{2}\equiv 
P^{\mu}P_{\mu}$ and  $W^{2}$, where  
$W_{\mu}=(-1/2)\epsilon_{\lambda\rho\nu\mu}
M^{\lambda\rho}P^{\nu}$ defines the  Pauli-Lubanski pseudovector. It follows
from $[W_{\mu},W_{\nu}]=i\epsilon_{\mu\nu\lambda\rho} W^{\lambda}P^{\rho}, 
\quad [W_{\mu},P_{\rho}]=0\;$  and  $\;W.P=0$ that in a   
representation charactarized by particualr eigenvalues of 
the two Casimir operators we 
may simultaneously diagonalize  $P^{\mu}$ along with just one 
component of  $W^{\mu}$. We have 
$ W^{+} =-[J_{3} P^{+}+B_{1} P^{2}-B_{2} P^{1}], 
W^{-} =J_{3} P^{-}+S_{1} P^{2}-S_{2} P^{1}, 
W^{1} =K_{3} P^{2}+ B_{2} P^{-}- S_{2} P^{+},$ and 
$W^{2} =-[K_{3} P^{1}+ B_{1} P^{-}- S_{1} P^{+}]$ and it shows 
that  $W^{+}$ has a  special place since it 
contains only the kinematical generators. On the LF we define  
  ${\cal J}_{3}= -W^{+}/P^{+}$ as the   {\it spin operator}$^{11}$. 
It may be shown  to 
commute with  $P_{\mu}, B_1,B_2,J_3,$ and  $K_3$. 
For $m\ne 0$ we may use the parametrizations $p^{\mu}:( p^{-}=(m^{2}+
{p^{\perp}}^{2})/(2p^{+}), 
p^{+}=(m/{\sqrt 2})e^{\omega}, 
p^{1}=-v_{1}p^{+},  p^{2}=-v_{2}p^{+})$ and 
${\tilde p}^{\mu}: (1,1,0,0)(m/{\sqrt 2})$ in the rest frame. We have 
$P^{2}(p)= m^{2} I$ and  $W(p)^{2}=
W(\tilde p)^{2}= -m^{2} [J_{1}^2+J_{2}^2+J_{3}^2] = -m^{2} s(s+1) I$ 
where $s$ assumes half-integer values.  
Starting from the rest  state $\vert \tilde p; m,s,\lambda, ..\rangle $
with ${J}_{3}\, \vert\tilde p; m,s,\lambda, ..\rangle
= \lambda \,\vert \tilde p; m,s, \lambda, ..\rangle $ we may build  an 
arbitrary eigenstate of $P^{+}, P^{\perp}, {\cal J}_{3} $  (and $ P^{-}$ ) 
on the LF by 
$$\vert p^{+},p^{\perp}; m,s,\lambda, ..\rangle= e^{i(\bar v. \bar B)} 
e^{-i\omega K_{3}} \vert \tilde p; m,s,\lambda, ..\rangle 
\eqno(1)$$ 

\nl If we make use of the following  {\it identity} for the  spin operator

$${\cal J}_{3}(p)=\;J_{3}+v_{1}B_{2}-v_{2}B_{1}=\quad 
e^{i(\bar v. \bar B)} \;
J_{3} \;e^{-i(\bar v. \bar B)}  \eqno(2)$$

\nl we find ${\cal J}_{3}\, \vert p^{+},p^{\perp}; m,s,\lambda, ..\rangle
= \lambda \,\vert p^{+},p^{\perp};m,s,\lambda, ..\rangle $. 
Introducing  also  ${\cal J}_{a}= 
-({\cal J}_{3}P^{a}+W^{a})/{\sqrt{P^{\mu}P_{\mu}}},$ $a=1,2$, which contain 
dynamical generators we verify that 
$\;[{\cal J}_{i},{\cal J}_{j}]=i\epsilon_{ijk} {\cal J}_{k}$.

For $m=0$ case when $p^{+}\ne0$ 
a convenient parametrization is 
$p^{\mu}:( p^{-}=p^{+} {v^{\perp}}^{2}/2, \,p^{+}, 
p^{1}=-v_{1}p^{+}, p^{2}=-v_{2}p^{+})$ and $\tilde p: 
(0, p^{+}, 0^{\perp})$. We have 
$W^{2}(\tilde p) = -(S_{1}^{2}+S_{2}^{2}){p^{+}}^{2}$ 
and  $[W_{1},W_{2}](\tilde p)=0, \,
[W^{+},W_{1}](\tilde p)=-ip^{+}W_{2}(\tilde p), 
\, [W^{+}, W_{2}](\tilde p)=ip^{+}W_{1}(\tilde p)$ showing that 
$W_{1}, W_{2}$ and  $W^{+}$  generate the algebra 
$SO(2)\otimes T_{2}$. The eigenvalues of 
$W^{2}$ are hence not quantized and they vary continuously. 
This is contrary to the experience so we impose that the physical states 
satisfy in addition  $W_{1,2}
\vert \, \tilde p;\,m=0,..\rangle=0$. Hence 
$ W_{\mu}=-\lambda P_{\mu}$ and  the invariant parameter 
$\lambda $ is taken to define as the {\it spin} of the massless particle.  
From  $-W^{+}(\tilde p)/{\tilde p}^{+}=J_{3}$ we conclude that 
$\lambda$ assumes half-integer values as well.  
We note that   $W^{\mu}W_{\mu}=\lambda^{2} P^{\mu}P_{\mu}=0$  and that 
on the LF the definition of the spin operator  appears  unified 
for massless and massive particles. A parallel discussion based on 
$p^{-}\ne0$ may also be given.

As an illustration consider the three particle state on the LF with the 
total eigenvalues 
$p^{+}$,  $\lambda$  and $p^{\perp}$. In the {\it standard frame } 
with  $p^{\perp}=0\;$ it may be written as 
($\vert x_{1}p^{+},k^{\perp}_{1}; \lambda_{1}\rangle
\vert x_{2}p^{+},k^{\perp}_{2}; \lambda_{2}\rangle
\vert x_{3}p^{+},k^{\perp}_{3}; \lambda_{3}\rangle$ ) 
with  $\sum_{i=1}^{3} \,x_{i}=1$, $\sum_{i=1}^{3}\,k^{\perp}_{i}=0$, and 
$\lambda=\sum_{i=1}^{3}\,\lambda_{i}$. Aplying 
$e^{-i{(\bar p.\bar B)/p^{+}}}$ on it we obtain 
($\vert x_{1}p^{+},k^{\perp}_{1}+x_{1}p^{\perp}; \lambda_{1}\rangle
\vert x_{2}p^{+},k^{\perp}_{2}+x_{2}p^{\perp}; \lambda_{2}\rangle
\vert x_{3}p^{+},k^{\perp}_{3}+x_{3}p^{\perp}; \lambda_{3}\rangle $ ) 
now with  $p^{\perp}\ne0$. The  $x_{i}$ and $k^{\perp}_{i}$ indicate 
relative (invariant) parameters and do not depend upon the 
reference frame. The  $x_{i}$ is 
the fraction of the total longitudinal momentum 
carried by the  $i^{th} $ particle while 
 $k^{\perp}_{i}$ its transverse momentum. The state of a pion with 
 momentum ($p^{+},p^{\perp}$),  for example, 
 may be expressed$^{}$ as 
an expansion over the LF Fock states constituted by the different 
number of partons  

$$\vert \pi : p^{+},p^{\perp} \rangle=
{\sum}_{n,\lambda}\int {\bar \Pi}_{i}{{dx_{i}d^{2}{k^{\perp}}_{i}}\over
{{\sqrt{x_{i}}\,16\pi^{3}}}} \vert n:\,x_{i}p^{+},x_{i}p^{\perp}+
{k^{\perp}}_{i},
\lambda_{i}\rangle\;\psi_{n/\pi}(x_{1},{k^{\perp}}_{1},
\lambda_{1}; x_{2},...) \eqno(3) $$

\nl where the summation is over all the Fock states 
$n$ and spin projections  $\lambda_{i}$, with 
${\bar\Pi}_{i}dx_{i}={\Pi}_{i} dx_{i}\; \delta(\sum x_{i}-1), $ 
and ${\bar\Pi}_{i}d^{2}k^{\perp}_{i}={\Pi}_{i} dk^{\perp}_{i} \;
 \delta^{2}(\sum k^{\perp}_{i})$.  The wave function of the 
parton $\psi_{n/\pi}(x,k^{\perp})$ 
indicates the probability amplitue for finding inside the pion 
the partons in the Fock state $n$ carrying 
the 3-momenta 
$(x_{i}p^{+}, x_{i}p^{\perp}+ k^{\perp}_{i}) $. The 
  Fock state of the pion is also 
	off the energy shell : $\,\sum k^{-}_{i} > p^{-}$.

The {\it discrete symmetry} transformations may also be defined on the 
LF Fock states. 
For example, under the conventional parity ${\cal P} $ 
the  spin operator  ${\cal J}_{3}$ is not left invariant. 
We may rectify this by defining {\it LF Parity operation} by 
${\cal P}^{lf}=e^{-i\pi J_{1}}{\cal P}$. We find 
then  $B_{1}\to -B_{1}, B_{2}\to B_{2}, P^{\pm}\to P^{\pm}, P^{1}\to 
-P^{1}, P^{2}\to P^{2}$ etc. such that 
${\cal P}^{lf}\vert p^{+},p^{\perp}; m,s,\lambda, ..\rangle
\simeq \vert p^{+},-p^{1}, p^{2}; m,s,\,-\lambda, ..\rangle $. Similar
considerations apply for charge conjugation and  time inversion. For example, 
it is straightforward to construct  the free {\it LF Dirac spinor}  
$\chi(p)= 
[\sqrt{2}p^{+}\Lambda^{+}+(m-\gamma^{a}p^{a})\,\Lambda^{-}]\tilde \chi/
{ {\sqrt{\sqrt {2}p^{+}m}}}$ which is also an eigenstate os 
${\cal J}_{3}$ with eigenvalues 
$\pm 1/2$. Here $\Lambda^{\pm}= \gamma^{0}\gamma^{\pm}/{\sqrt 2}=
\gamma^{\mp}\gamma^{\pm}/2=({\Lambda^{\pm}})^{\dagger}$, 
$ (\Lambda^{\pm})^{2}=\Lambda^{\pm}$,  
and $\chi(\tilde p)\equiv \tilde \chi\,$ with  $\gamma^{0}
\tilde \chi= \tilde \chi$. The conventional (equal-time) 
spinor can also be constructed by the  procedure analogous to that 
followed for the LF spinor and it has  the well known form 
$ \chi_{con}(p)=  (m+\gamma.p)\tilde \chi/
{\sqrt{2m(p^{0}+m)}}$. 
Under the conventional parity operation  ${\cal P}: 
\chi'(p')=c \gamma^{0} \chi(p)$ ( since we must require 
$\gamma^{\mu}={L^{\mu}}_{\nu}\,S(L)\gamma^{\nu}{S^{-1}}(L)$ etc. ). We find 
$\chi'(p)=c 
[\sqrt{2}p^{-}\Lambda^{-}+(m-\gamma^{a}p^{a})\,\Lambda^{+}]\,\tilde \chi
/{\sqrt{\sqrt {2}p^{-}m}}$. For $ p \neq\tilde p$ 
it is not proportional to $\chi(p)$ in contrast to the result in 
the case of the usual spinor where 
$\gamma^{0}\chi_{con}(p^{0},-\vec p)=\chi_{con}(p)$ for  $E>0$ (and  
$\gamma^{0}\eta_{con}(p^{0},-\vec p)=-\eta_{con}(p)$ for  $E<0$). 
However, applying parity operator twice we do show 
$\chi''(p)=c^{2}\chi(p)$ hence leading to the usual result 
$c^{2}=\pm 1$. The LF parity operator over spin $1/2$ Dirac spinor is 
${\cal P}^{lf}= c \,(2J_{1})\,\gamma^{0}$ and the corresponding transform 
of $\chi$ is shown to be an  eigenstate of ${\cal J}_{3}$. 

\bigskip

\nl {\bf 4. \quad Spontaneous Symmetry Breaking ({\rm SSB}) Mechanism.  
Continuum Limit \qquad of Discretized LF Quantized Theory. 
Nonlocality of LF Hamiltonian }
\medskip

 The quantization of  scalar theory  in equal-time framework 
 is found in the text books but the  
existence of the  continuum limit of the Discretized Light Cone Quantized 
(DLCQ)$^{12}$ theory, the nonlocal nature of the LF Hamiltonian, and the 
description of the SSB on the LF were 
clarified$^{}$ only recently. 

 Consider first the two 
dimensional case  with 
${\cal L}= \; \lbrack
{\dot\phi}{\phi^\prime}-V(\phi)\rbrack$. 
 Here $\tau\equiv x^{+}=(x^0+x^1)/{\sqrt2}$,
  $x\equiv x^{-}=(x^0-x^1)/{\sqrt2}$, $\partial_{\tau}\phi=\dot\phi , 
	\partial_{x}\phi={\phi}'$, and $d^2x=d\tau dx$. The eq. of motion, 
$\,\dot{\phi^\prime}=(-1/2)\delta V(\phi)/\delta \phi $, shows that 
$\phi=const. $ is a possible solution. 
We write$^{13}$ 
$\;\phi(x,\tau)=\omega(\tau)+\varphi(x,\tau)\;$ where $\omega(\tau)$ 
corresponds to the {\it bosonic condensate} and $\varphi(\tau,x)$ describes 
(quantum) {\it fluctuations} above it. 
The value of   $\omega(=\langle 0\vert \phi\vert 0\rangle)$
will be seen to characterize the corresponding vacuum  state. 
The translational  invariance of the ground state requires 
that $\omega $ be a constant    so that 
${\cal L}={\dot\varphi}{\varphi}^\prime-V(\phi)$.  
Dirac procedure$^{9}$ is applied now  to 
construct  Hamiltonian theory which would permit$^{1}$ us to 
to construct a 
quantized relativistic field theory. We may avoid using 
distribuitions if we   restrict $x$ to a finite interval from 
 $\,-L/2\,$ to $\,L/2\,$. 
 The 
 {\it physical  limit$^{}$ to the continuum ($L\to\infty\,$)}, however, 
  must be taken latter to remove the spurious finite volume effects. 
Expanding $\varphi$ by Fourier series we obtain 
	$\phi(\tau,x) \equiv \omega +\varphi(\tau,x)
= \omega+{1\over\sqrt{L}} {q_{0}(\tau)}+
{1\over\sqrt{L}}\;{{\sum}'_{n\ne 0}}\;\;
{q}_n(\tau)\;e^{-ik_n x}$ 
 where $\,k_n=n(2\pi/L)$, 
$\,n=0,\pm 1,\pm 2, ...\,$ and 
the {\it discretized theory} Lagrangian becomes 
$\;i{{\sum}_{n}}\;k_n \,{q}_{-n}\;{{\dot q}}_{n}-
                                  \int dx \; V(\phi)$. 
The momenta conjugate to ${q}_n$ are 
 ${p}_n=ik_n{q}_{-n}$ 
 and the canonical LF Hamiltonian is found to be  $\int \,
 dx \,V(\omega+\varphi(\tau,x))$. 
The primary constraints$^{}$ are thus 
${p}_0 \approx 0\;$ and $\;{\Phi}_n \equiv 
\;{p}_n-ik_n{q}_{-n}\approx 0\;$ for $\,n\ne 0\,$. 
We follow$^{14}$ the standard  Dirac procedure$^{9}$ and find 
three  {\it weak constraints}$^{9}$  $\,p_{0}\approx 0$, $\,\beta
\equiv \int dx \,V'(\phi) \approx 0$, and $\,\Phi_{n}\approx 0\,$ 
for $\,n\ne 0\,$ 
on the phase space and they are shown to be {\it second class}$^{9}$. 
We find $(\,n,m\ne 0\,)$
$\{{\Phi}_n, {p}_0\}\,=\,0,\qquad 
\{{\Phi}_n,{\Phi}_m\}\,=\,-2ik_n \delta_{m+n,0}\;,$
$\quad \{{\Phi}_n,\beta\}\,=\,\{{p}_n,\beta\}\,
=\,-{(1/ \sqrt{L})}\int dx\;\lbrack \,V''(\phi)-V''([{\omega +
q_{0}]/
\sqrt{L}})\,\rbrack \,e^{-ik_nx}\,
\equiv \,-{{\alpha}_{n}/\sqrt{L}},\,$
\quad $\{{p}_0,\beta\}\,=\,-{(1/\sqrt{L})}\int dx\;V''(\phi)\,
\equiv \,-{\alpha/\sqrt{L}},\,$\quad
$\{{p}_0,{p}_0\}\,=\,\{\beta,\beta\,\}\,=\,0\,$. 
Implement first the pair of constraints $\;{p}_0\,\approx 0, 
\,\beta\,\approx 0$ by modifying the Poisson brackets to 
the star  bracket $\{\}^*$  defined by 
$\{f,g\}^*\,=\{f,g\}\,-\lbrack\,\{f,{p}_0\}\;
\{\beta,g\}-\,(p_0\,\leftrightarrow 
\beta)\rbrack\,({\alpha/\sqrt{L}})^{-1}$.
We may then set ${p}_0\,=0$ 
and $\beta \,=0$ as {\it strong relations}$^{9}$. 
We find  by inspection that the brackets 
$\{\}^*\,$ of the remaining  variables coincide with the standard 
Poisson brackets except for the ones involving $q_{0}$ and 
$\,p_{n}\,$ ($n\ne0$):\quad 
$\,\{q_0,{p}_n\}^*\,=
\{q_0,{\Phi}_n\}^*\,=-({\alpha^{-1}}{\alpha}_n)\,$.  
For example, if 
$V(\phi)=\,({\lambda/4})\,{(\phi^2-{m^2}/\lambda)}^2\;$, 
$\lambda\ge 0, m\ne 0\,$ 
we find  $\;\{q_0,{p}_n\}^*\;
[\{\,3\lambda\,({\omega+q_{0}/\sqrt{L}})^{2}-m^{2}\,\}L\,+
6\lambda(\omega+q_{0}/{\sqrt L})\int\, dx \varphi +
\,3\lambda \,\int\,dx\,\varphi^{2}\,]= 
-\,3\lambda\,[\,2(\omega+q_{0}/{\sqrt L})\,{\sqrt L}
 q_{-n}
 + \int \,dx\,\varphi^{2} 
\, e^{ -ik_{n}x} \,] $.

We next implement the constraints $\,\Phi_{n}\approx 0\,$ 
($\,n\ne 0$). We have $ C_{nm}\,=\,\{\Phi_{n},\Phi_{m}\}^*\,=
\,-2ik_{n}\delta_{n+m,0}\,$ 
and its inverse is given by  $\,{C^{-1}}_{nm}\,=\,(1/{2ik_{n}})
\delta_{n+m,0}\;$. The {\it final} Dirac bracket which taking care of all 
the constraints is then given by

$$\{f,g\}_{D}\,= \,\{f,g\}^{*}\,-\,{{\sum}'_{n}}\,{1\over {2ik_{n}}}
\{f,\,\Phi_{n}\}^*\,\{\Phi_{-n},\,g\}^*.\,\eqno(4)$$

\noindent where we may now in addition write $\,p_{n}\,=\,ik_{n}
q_{-n}\,$. It is easily shown that 
$\{q_0,{q}_0\}_{D}\,=0, 
\{q_0,{p}_n\}_{D}\,=\{q_0,\,ik_{n}
q_{-n}\}_{D}\,={1\over 2}\,\{q_0,{p}_n\}^*,
\{q_{n},p_{m}\}_{D}\,={1\over 2}\delta_{nm}$. 

The  limit to the continuum$^{14}$, $L\to\infty$ is taken as usual:  
$\Delta=2\,({\pi/{L}})\to dk\,,\,k_{n}=n\Delta\to k\,,\,\sqrt{L}\,
q_{-n}\to\,lim_{L\to\infty}
 \int_{-L/2}^{L/2}{dx}\,\varphi(x)\,e^{ik_{n}x}\equiv\,\int_{-\infty}^
{\infty}\,dx\, \varphi(x)\,e^{ikx}\,=\,\sqrt{2\pi}{\tilde\varphi(k)}$ 
for all $\,n $,  
$\,\sqrt{2\pi}
\varphi(x)\,=\int_{-\infty}^{\infty}\,dk\,\tilde\varphi(k)\,e^{-ikx}\;$, 
and  $(q_{0}/{\sqrt{L}})\to 0\,$. 
From $\,\{\sqrt{L}q_{m},\sqrt{L}q_{-n}\}_{D}\,=\,
L\,\delta_{nm}/({2ik_{n}})\,$ following from $\{q_{n},p_{m}\}_{D}$ 
for $n,m\ne 0$  
we derive, on using  $\,L\delta_{nm}\to \int_{-\infty}^{\infty}dx e^{i(k-k')x}=
\,{2\pi\delta(k-k')}$, that 
$\{\tilde\varphi(k),\tilde\varphi(-k')\}_{D}\,=\,\delta
(k-k')/(2ik)\,$ 
where  $ k,k'\,\ne 0$. 
If we use the integral representation of the sgn 
function  the well known LF Dirac bracket 
$\{\varphi(x,\tau),\varphi(y,\tau)\}_{D}=-{1\over 4}\epsilon(x-y)$ is
obtained. 
The expressions of  $\{q_{0},p_{n}\}_{D}$ (or $\{q_{0},\varphi'\}_{D}$) 
show that the DLCQ is harder to work with here. 
The  continuum limit of  $\beta=0$ is 

$$ \eqalign {& \,lim_{L\to\infty} 
{1\over L}\int_{L/2}^{L/2} dx \,V'(\phi)\equiv  \cr 
& \omega(\lambda\omega^2-m^2)+lim_{L\to\infty} {1\over L}
 \int_{-L/2}^{L/2} dx \Bigl[ \,(3\lambda\omega^2-m^2)\varphi + 
 \lambda (3\omega\varphi^2+\varphi^3 ) \,
\Bigr]=0}\,\eqno(5)$$

\nl while that for the LF Hamiltonian is ($P^{-}\equiv H^{l.f.}$)

$${ P^{-}\,=\int  dx \,\Bigl [\omega(\lambda\omega^2-m^2)\varphi+
{1\over 2}(3\lambda\omega^2-m^2)\varphi^2+
\lambda\omega\varphi^3+{\lambda\over 4}\varphi^4 
\Bigr ]\,}\eqno(6)$$

\nl These results  follow  immediately 
if we worked directly  in the continuum formulation$^{13}$; we do have 
to handle generalized functions now.  In the LF Hamiltonian theory 
we have an additional new ingredient in the form of the 
{\it constraint equation} (5).  Elimination of 
$\omega $ using it would lead to a {\it nonlocal LF Hamiltonian}$^{14}$   
in contrast to   the corresponding 
local one in the equal-time formulation.  
At the tree level the integrals appearing in (5) 
are convergent from the theory of Fourier transform. 
When  $L\to\infty$, it results in  $V'(\omega)=0$, which in the
equal-time theory is essentially {\it added} to  it 
as an  external constraint. In 
the renormalized theory$^{15}$  the constraint equation 
 describes 
the high order quantum corrections to the tree level value of the condensate.

The quantization is performed via the correspondence$^{}$  
$i\{f,g\}_{D}\to [f,g]$. Hence 
$\varphi(x,\tau)= {(1/{\sqrt{2\pi}})}\int dk\; 
{\theta(k)}\;
[a(k,\tau)$ $e^{-ikx}+{a^{\dag}}(k,\tau)e^{ikx}]/(\sqrt {2k})$, 
were $a(k,\tau)$ and ${a^{\dag}}(k,\tau)$ 
satisfy the canonical equal-$\tau$ commutation relations, 
$[a(k,\tau),{a(k^\prime,\tau)}^{\dag}]=\delta(k-k^\prime)$ etc.. 
The vacuum state is defined 
by  $\,a(k,\tau){\vert vac\rangle}=0\,$, 
$k> 0$ and the tree level  
description of the {\it SSB} is  given as
follows. The values of $\omega=
\,{\langle\vert \phi\vert\rangle}_{vac}\;$ obtained from  
$V'(\omega)=0$ 
{\it characterize} the different   vacua in the theory. 
Distinct  Fock spaces corresponding to different values of $\omega$ 
are built as usual by applying the creation operators on the corresponding
vacuum state.  The $\omega=0$ corresponds to a {\it symmetric phase} 
since the hamiltonian is then symmetric under 
$\varphi\to -\varphi$. For $\omega\ne0$ this symmetry is 
violated and the  system is 
in a {\it broken or asymmetric phase}.  

The self-consistency$^{9}$ may also be checked. Hamilton's eq. gives 
$\dot\varphi(x,\tau)= -i\, [\varphi(x,\tau),H^{l.f.}(\tau)]
= -\int dy \,\epsilon(x-y)\, V'(\phi(y,\tau))/4\, $ and  
we recover the Lagrange eq. $\dot\varphi'(x,\tau)=- V'( \phi(x,\tau))/2$. 
If we substitute the value of $V'(\phi)$ obtained 
from the latter in the former  we find after an integration by parts 
$ \dot\varphi(x,\tau)=\dot\varphi(x,\tau)-
{ \Bigl [\dot\varphi(\infty,\tau)\epsilon(\infty-x)
-\dot\varphi(-\infty,\tau)\epsilon(-\infty-x)\Bigr]/2}$. 
For finite values of $x$ this leads to 
$\,\dot\varphi(\infty,\tau)+\dot\varphi(-\infty,\tau)=0\,$. On the 
other hand, if we integrate  the momentum space expansion 
of $\varphi'(x,\tau)$ given above  
we  may show that $\,\varphi(\infty,\tau)-\varphi(-\infty,\tau)=0$. 
Hence we are led to 
$\partial_{\tau}\varphi(\pm\infty,\tau)=0$ as a self-consistency condition. 
This is analogous to the condition 
$\partial_{t}\varphi(x^{1}=\pm\infty,t)=0$ which 
in contrast is {\it added} to  the equal-time theory upon invoking 
physical considerations. The constraint eq. is then seen to 
follow  also upon 
a  space  integration of the Lagrange eq.. 
A self-consistent 
Hamiltonian formulation can thus be built in the 
continuum which can also describe the {\it SSB}. 

The extension$^{14}$ to $3+1$ dimensions and to global continuous symmetry 
is straightforward. Consider  real scalar fields 
$\phi_{a} (a=1,2,..N)\,$ which form  an isovector of global 
internal symmetry group 
 $O(N)$. We now  write 
 $\phi_{a}(x,\bar x,\tau)
=\omega_{a}+\varphi_{a}(x,\bar x,\tau)$ and 
the Lagrangian density is  ${\cal L}=[{\dot\varphi_{a}}{\varphi'_{a}}-
{(1/ 2)}(\partial_i\varphi_{a})(\partial_i\varphi_{a})-V(\phi)]$,  
where $i=1,2$ indicate the transverse space directions. The Taylor series 
expansion of the constraint equations $\beta_{a}=0$ gives a set of coupled
eqs. $L\,V'_a(\omega)+
\,V''_{ab}(\omega)\int dx \varphi_{b}+\,
V'''_{abc}(\omega)\int dx \varphi_b\varphi_c/2+...=0$. Its discussion at 
the tree level leads to the conventional theory results. 
The  LF symmetry generators are found to be 
$G_{\alpha}(\tau)=
-i\int d^{2} {\bar x} dx \varphi'_{c}(t_{\alpha})_{cd}\varphi_{d} 
=\,\int d^{2}{\bar k}\,dk \, \theta(k) {a_{c}(k,{\bar k})
^{\dag}} (t_{\alpha})_{cd} a_{d}(k,{\bar k}) $ where 
$\alpha,\beta=1,2,..,N(N-1)/2\, $,  are the group indices, 
$t_{\alpha}$ are hermitian and antisymmetric generators of $O(N)$, and 
${a_{c}(k,{\bar k})^{\dag}} ( a_{c}(k,{\bar k}))$ is creation (
destruction)  operator  contained in the momentum space expansion 
of $\varphi_{c}$. These  are to be contrasted with the generators 
in the equal-time theory,  
$ Q_{\alpha}(x^{0})=\int d^{3}x \, J^{0}
=-i\int d^{3}x (\partial_{0}\varphi_{a})(t_{\alpha})_{ab}\varphi_{b} 
-i(t_{\alpha}\omega)_{a}\int d^{3}x 
({{d\varphi_{a}}/ dx_{0}})$.  
Thus the  generators on the LF   always 
annihilate the LF vacuum  and the SSB is  
now seen  in the 
broken symmetry of the quantized theory Hamiltonian. The criterian for 
the counting of the number of Goldstone bosons on the LF follows 
to be the same as in the conventional theory. 
On the other hand, the first term 
on the right hand side of $Q_{\alpha}(x^{0})$ does  annihilate  
the conventional theory vacuum  but 
the second term  gives now  non-vanishing contributions  
for some of the (broken) generators. The  symmetry of the vacuum 
is thereby broken  while the quantum 
Hamiltonian remains invariant.  
The physical content  of SSB in the {\it instant form}
and the {\it front form}, however, is the same 
though achieved  by differnt descriptions. Alternative proofs$^{14}$ 
 on the LF,  
in two dimensions, can be given  
of the Coleman's theorem related to the absence of 
Goldstone bosons and  of the pathological nature of 
massless scalar theory; we are unable to implement 
the second class constraints over the phase space.

We remark that the simplicity of the LF vacuum is in a sense 
compensated by the involved nonlocal Hamiltonian. The latter, however, 
may be treatable using advance computational techniques. In a recent 
work$^{15}$ it was also shown that renormalized theory may be constructed 
without the need of first solving the constraint eq.  for $\omega$.  
Instead we  perform  renormalization and 
obtain a renormalized constraint equation. For 
$(\phi^{4})_{2}$ theory  
this along with the equation expressing  mass renormalization condition are
sufficient to describe the phase transition in the theory. It was 
found to be of the second order, which agrees with the conjecture of 
Simon and Griffiths$^{16}$, in contrast to the first order transition 
found if we follow the variational methods.

\baselineskip=18pt
\bigskip
\nl {\bf 5. \quad Chern-Simons ({\rm CS}) Gauge Theory }
\medskip 

LF quantization may turn out to be useful for nonperturbative 
computations in QCD  and in the study of  relativistic bound states 
of light fermions. To elucidate some  general  features in  gauge theory 
quantized on the LF we consider$^{17}$ the  CS  theory described by the 
singular Lagrangian 
${{\cal L}} =({{\cal D}}^{\mu}\phi)({\widetilde{\cal D}}_{\mu}\phi^{*})\,+
({\kappa/ {4\pi}}){\epsilon}^{\mu\nu\rho} 
A_{\mu} \partial_{\nu} A_{\rho}$,  which is known to be 
relevant for the theory 
of {\it anyons}-  excitations with fractional statistics. 
Here ${{\cal D}}_{\mu}= (\partial_{\mu}+ie A_{\mu})$, 
${\widetilde {\cal D}}_{\mu}= (\partial_{\mu}-ie A_{\mu})$,  and 
the theory a has a conserved and gauge invariant four-vector 
current $j^{\mu}=ie(\phi^{*}{{\cal D}}^{\mu}\phi-
\phi{\widetilde{\cal D}}^{\mu}\phi^{*})\, $. 
Its contravariant vector property  must remain  intact  
if the Hamiltonian theory constructed is relativistic. 

On the LF the light cone gauge (l.c.g.),  $A_{-}=0$, is clearly 
accessible in the Lagangian formulation. It will  be shown to 
be so also  on the phase space. Before applying the Dirac method 
to construct an Hamiltonian 
we must consider the boundary conditions (bcs) on the fields involved in 
our non-covariant gauge. 
The self-consistency$^{9}$ requires that 
the Hamiltonian theory  must not contradict the Lagrangian theory 
and we may thus examine first the Lagrange eqs. in l.c.g.. We find 
an expression of the electric charge $Q$ 
on integrating (one of) the eq. of motion $2a\partial_{-}A_{1}=j^{+}$, 
where  ${\kappa= {4 \pi}a}$:  
$Q=\int d^2x \,j^{+}=2a \int   
dx^{1}\,[A_{1}(x^{-}=\infty,x^{1})-A_{1}(x^{-}=-\infty,x^{1})]\,$. 
It follows that if the  charge is nonvanishing  
 $A_{1}$ can not 
satisfy the periodic or the vanishing  bcs 
at infinity  along  $x^{-}$. 
We will assume the  {\it anti-periodic} bcs 
for the gauge fields  
along $x^{-}$ and  the vanishing ones along  
 $x^{1}$.  For the scalar fields similar arguments allow us to 
assume vanishing  bcs at infinity. 
The canonical Hamiltonian, after integration by parts using these 
bcs,  may then be written as 
$ H_{c}=\int d^2x \bigl[\,({\cal D}_{1}\phi)
({\widetilde {\cal D}}_{1}\phi^{*})-A_{+} \Omega \bigr] $
where $\Omega=ie(\pi\phi-\pi^{*}\phi^{*})+a \epsilon^{+ij}\partial_{i}A_{j}
+\partial_{i}\pi^{i}$ and $i=-,1$. From this as the  starting point$^{17}$  
we apply the  Dirac procedure$^{9}$ to construct a self-consistent 
Hamiltonian theory corresponding to the singular CS Lagrangian. We find 
two first class constraints $\pi^{+}\approx 0$ and 
$\Omega \approx 0$ which generate gauge transformations 
and four second class ones, 
$\top\equiv \pi - {\widetilde {\cal D}}_{-} \phi^{*}  \approx 0, 
\top^{*}\equiv \pi^{*}- {\cal D}_{-}\phi \approx 0 $, and 
$ \top^{i}\equiv\pi^{i}-a \epsilon^{+ij} A_{j}  \approx 0 $. 
The extended Hamiltonian is 
$H'= H_{c}+ \int d^2x \bigl [ \, u\top+u^{*}\top^{*}+u_{i}\top^{i}+
u_{+}\pi^{+}\, \bigr ]
$  where $u,u^{*},u^{i},u_{+} $, (and $A_{+}$) are Lagrange multiplier fields. 
The  eqs. of motion are obtained from $df(x,\tau)/d\tau\,= 
\{f(x,\tau),H'(\tau)\}+\partial f/\partial \tau$ and from them 
we conclude that a set of multipliers may be chosen such that 
$A_{-}\approx 0$ and $dA_{-}/d\tau\approx 0$. The  {\it local} l.c.g. 
$A_{-}\approx 0$ is thus also accessible on the phase space. We add 
in the theory this gauge-fixing constraints so that now 
the set of second class constraints becomes 
$\top_{m},\quad m=1,2..6$: 
$\top_{1}\equiv\top^{-},\top_{2}\equiv\top^{1},\top_{3}\equiv \top, 
\top_{4}\equiv \top^{*}, \top_{5}\equiv A_{-}, \top_{6}\equiv \Omega\,$ 
while $\pi^{+}\approx0$ stays first class.  
The  initial Poisson brackets are now modified to 
define  the Dirac brackets  $\{f,g\}_{D}$ such that 
the second class constraints may be written as {\it strong equalities}$^{9}$ 
 $\top_{m}=0$ and 
$df(x,\tau)/d\tau\,= \{f(x,\tau),H'(\tau)\}_{D}+\partial f/
\partial \tau$. The Dirac brackets are constructed$^{17}$ to be 

$$\{f,g\}_{D}= \{f,g\}-\int d^2u d^2v \, \{f,\top_{m}(u)\}\,
C^{-1}_{mn}(u,v)\,\{\top_{n}(v),g \} \eqno(7) $$

\nl where $\, C^{-1}(x,y)\,$ is  given by 

$$ {\left (\matrix {
0      & -4 a{{\partial}^{x}}_{-} & 0   & 0   & 0   & 0  \cr
4 a {{\partial}^{x}}_{-}     &  
[\phi^{*}(x)\phi(y)+\phi(x)\phi^{*}(y)]
   & {2ai}\phi(x)
    & - {2ai}\phi(x)^{*}      &  0    &  -{4 a}  \cr
0   &   {2ai}\phi(y)  & 0      &  (2a)^2 & 0  & 0  \cr
0  & - {2ai}\phi^{*}(y)  & (2a)^2  & 0 & 0 &  0 \cr
0  & 0 & 0  &  0 & 0   &  2(2a)^2  \cr
0 & -{4 a} & 0 & 0 & 2(2a)^2  & 0 \cr } \right)} {K(x-y)\over ({2a})^2}
\eqno(8) $$

\medskip
\nl It is the inverse of the constraint matrix 
with the elements $C_{mn}=\{\top_{m},\top_{n}\}$ and  
 $\,K(x-y)=-(1/4)\,\epsilon(x^{-}-y^{-})\,\delta(x^{1}-y^{1})$. 
We find that $A_{+}$ which is already 
absent in $\top_{m}$, drops out also from $ H_{c}$ since $\Omega=0$. 
The $\pi^{+}\approx 0$ 
stays first class even with respect to the Dirac brackets and the 
multiplier  $u_{+}$ is left undetermined. The variable $\pi^{+}$ 
decouples and we may choose $u_{+}=0$ so that $\pi^{+}$ and 
$A_{+}$ are eliminated. 
The  LF  Hamiltonian then simplifies to

$$H^{l.f.}(\tau)=
\int d^{2} x \, \,({\cal D}_{1}\phi)({\widetilde {\cal D}}_{1}\phi^{*})
\eqno(9)$$

\nl There is still a $U(1)$ {\it global} 
gauge symmetry generated by $Q$. 
The scalar fields transform under this symmetry but they are left 
invariant under the local gauge transformations since, 
${ \{\Omega,f\}}_{D}=0 $. 
The only {\it independent variables} left are $\phi$ and $\phi^{*}$ 
which satisfy the well known equal-$\tau$ {\it LF  Dirac brackets}

$$ \{\phi,\phi\}_{D}=0, \,\, \{\phi^{*},\phi^{*}\}_{D}=0,
\quad \{\phi(x,\tau),\phi^{*}(y,\tau)\}_{D}= K(x,y)\quad \eqno(10)$$

We remark that we could alternatively eliminate $\pi^{+}$ by 
introducing {\it another local gauge-fixing weak condition} 
$A_{+}\approx 0$ (and $dA_{+}/d\tau\approx 0$) 
which is easily shown to be accessible. The  
additional modification of brackets does not alter the 
Dirac brackets of the scalar field already obtained. There is thus {\it no 
inconsistency in choosing the two 
{\sl local} and  {\sl weak} gauge-fixing conditions  $A_{\pm}\approx 0$ 
on the phase space  {\sl at one fixed  time $\tau$} 
in the CS gauge theory}; that they are  accessibile follows from 
the  Hamilton's eqs. of motion. 

We check now  the {\it self-consistency}. 
From the Hamilton's eq. for $\phi$ we derive 
($e=1, \pi^{*}=\partial_{-}\phi ):\;\;
 \partial_{-} \partial_{+}\phi(x,\tau)
 =\{\pi^{*}(x,\tau),H(\tau)\}_{D}= {1\over2}{{\cal D}}_{1}{{\cal D}}_{1}
\phi -i{\cal A}_{+}\partial_{-}\phi-{i\over 2} 
(\partial_{-}{\cal A}_{+})\phi $ where 
$\,-{2a\,}{\partial_{-}}{{\cal A}_{+}}={j^{1}}=
-ie(\phi^{*}{\cal D}_{1}\phi-
\phi {{\widetilde{\cal D}}_{1}}\phi^{*})$. 
On comparing  this   with the corresponding Lagrange eq.  
$\partial_{+}\partial_{-}\phi =
{1\over2}{{\cal D}}_{1}{{\cal D}}_{1}
\phi -i A_{+}\partial_{-}\phi-{i\over 2} (\partial_{-}A_{+})\phi $
in the l.c.g. it is suggested for convenience to 
  rename the expression 
${\cal A}_{+}$ on the phase space  by (the above eliminated ) 
$A_{+}$. We thus obtain agreement also with the other Lagrange eq. 
$\,-{2a\,}{\partial}_{-} A_{+}= j^{1} =-ie(\phi^{*}{{\cal D}_{1}}{\phi}-
{\phi}{\widetilde{\cal D}}_{1}{\phi^{*}})$. 
The Gauss' law eq. is seen to correspond to 
$\Omega=0$ and the remaining Lagrange eq. is also shown to be recovered. 
 The Hamiltonian theory in the l.c.g.  constructed here 
is thus shown  self-consistent. 
The variable $A_{+}$ has  
{\it reappeared} on the phase space and we have {\it effectively} 
 $A_{-}=0 $ (and {\it not} $A_{\pm}=0$). 
Similar discussion can be made in the Coulomb gauge in relation 
to $A^{0}$ and  there is 
{\it no inconsistency  on using the  
non-covariant local gauges} for the CS system.  
That 
only the nonlocal gauges$^{18}$ may describe consistently the excitations with 
fractional statistics 
 in the CS system does not agree with our conclusions. We find that  it should 
also arise 
in the quantum dynamics of the simpler Hamiltonian theory described 
by (9) and (10) on the LF  in 
the {\sl local l.c.g.}, which possibly may be used to construct renormalized theory 
of {\it anyons},   
or in the {\sl local} Coulomb gauge in the conventional framework. 
In the latter case or in the nonlocal gauges 
the  Hamiltonian  is complicated and renormalized 
theory seems difficult to construct.  A {\it dual description}$^{17,19}$ 
may also be constructed on the LF. 
We can rewrite the 
Hamiltonian density  as $ {\cal H}=(\partial_{1}\hat\phi) 
(\partial_{1}\hat \phi^{*})$ if we use  $A_{1}=\partial_{1}\Lambda$ 
where 
$\,8a\,\Lambda(x^{-},x^{1})=\int d^2y \,\epsilon(x^{-}-y^{-})\,
\epsilon(x^{1}-y^{1})\, 
j^{+}(y)\,$ and define $\,{\hat\phi}=e^{i\Lambda}\phi\,$,
 $\,{\hat\phi^{*}}=e^{-i\Lambda}\phi^{*}\,$. 
 The field $\hat\phi$ clearly does not have the vanishing Dirac
bracket (or commutator) with itself and leads to  manifest 
fractional statistics. 

The 
relativistic invariance of the theory above 
is shown$^{17}$ by checking the Poincar\'e algebra of the field theory 
space time symmetry generators.  We also come to the conclusion that 
the {\it anyonicity } 
seems not to be related to the unusual 
(not unexpected$^{17}$ in non-covariant gauges) 
behavior 
 under space rotations (sometimes referred to as rotational anomaly$^{20,19}$)  
of the scalar or the gauge field 
 but rather to the (renormalized) quantum dynamics of 
 CS system, for example,  described by (9) and (10).

\baselineskip=18pt
\bigskip
\nl {\bf 6. \quad  Conclusions }
\medskip 

The LF quantization seems useful and complementary to the conventional one 
and may be used with some advantage in the context of gauge theories like QCD 
and CS systems among others for stdying  nonperturbative effects. 
The description of the physical observation (like the SSB, Higgs 
mechanism, Anyonicity, Phase transition etc.) on the LF may be somewhat
different. The self-consistency conditions contained 
in the constrained dynamical system 
on the  LF (phase space) seem to correspond to 
(at least some of) the  external 
constraints we generally  add in the  conventional quantization 
on the basis of  physical considerations. The local non-covariant 
gauges$^{21}$ which have been successfully used in Yang-Mills  gauge theories 
may be used consistently also in the case of CS gauge theory.

\bigskip
\nl {\bf Acknowledgements}
\medskip
Acknowledgements are due to Ken Wilson, A. Harindranath, R. Perry and
M. Tonin 
for  constructive discussions and suggestions and to  
J. Leite Lopes,  F. Caruso, B. Pimentel and R. Shellard  for their comments.

\bigskip
\nl {\bf Appendix : \quad Commutators for equal-$x^{-}$}
\medskip
The LF formulation is symmetrical with respect to $x^{+}$ and $x^{-}$ 
and it is a matter of convention that we take the plus component as
the LF {\sl time} while the other as a spatial coordinate. 
The  theory quantized at $x^{+}=const.$  hyperplanes, however, does 
seem to already incorporate in it the information from 
the equal-$x^{-}$ quantized theory. 

For illustration we consider the two dimensional massive free scalar
theory. The LF quantization, assuming 
$x^{+}$ as the LF {\sl time} coordinates, leads  to 
$\omega=0$ and the equal-${x^{+}}$ commutator 
$[\varphi(x^{+},x^{-}),
\varphi(x^{+},y^{-}]=-i\epsilon(x^{-}-y^{-})/4$. The commutator 
can be realized in the momentum space through the expansion

$$\varphi(x^{+},x^{-})={1\over{\sqrt{2\pi}}}\int_{k^{+}>0}^{\infty}\,
 {dk^{+}\over{\sqrt{2k^{+}}}}
\Bigl [ a(k^{+}) e^{-i(k^{+}x^{-}+k^{-}x^{+})} +
        a^{\dag}(k^{+}) e^{i(k^{+}x^{-}+k^{-}x^{+})} \Bigr ]$$

\nl where $[a(k^{+}),{a(l^{+})}^{\dag}]=\delta(k^{+}-l^{+})$ etc. and
$2k^{+}k^{-}=m^{2}$.  It is then easy to show 

$$[\varphi(x^{+},x^{-}),
\varphi(y^{+},x^{-})]={1\over{2\pi}}\int_{k^{+}>0}^{\infty}
{dk^{+}\over {2k^{+}}}\Bigl [e^{ik^{-}(y^{+}-x^{+})}-
e^{-ik^{-}(y^{+}-x^{+})} \Bigr ].$$

\nl We may change the integration variable to $k^{-}$ by 
making use of $k^{-}dk^{+}+k^{+}dk^{-}=0$. Hence on employing the
integral representation 
$\epsilon(x)= (i/\pi){\cal P}\int_{-\infty}^{\infty} (d\lambda/\lambda) \;
exp(-i\lambda x)\,$ we arrive at the equal-$x^{-}$ commutator 

$$[\varphi(x^{+},x^{-}),
\varphi(y^{+},x^{-})]= -{i\over 4} \epsilon(x^{+}-y^{+})$$

\nl The above field expansion on the LF, in 
contrast to the equal-time case, does not 
involve the mass parameter $m$ and 
the same result follows in the massless case also  if we 
assume that $k^{+}=l^{+}$ implies $k^{-}=l^{-}$. 
Defining  the right  and the left movers  by 
$\varphi(0,x^{-})\equiv \varphi^{R}(x^{-})$, 
and $\varphi(x^{+},0)\equiv \varphi^{L}(x^{+})$ we obtain  
$[\varphi^{R}(x^{-}),\varphi^{R}(y^{-})]=(-i/4)\epsilon(x^{-}-y^{-})$ while 
$[\varphi^{L}(x^{+}),\varphi^{L}(y^{+})]=
(-i/4)\epsilon(x^{+}-y^{+})$.  

\bigskip

\bigskip
\nl {\bf References}
\medskip 

\item{1.}P.A.M. Dirac, Rev. Mod. Phys. {\bf 21} (1949) 392.
\item{2.}S. Weinberg, Phys. Rev. {\bf 150} (1966) 1313. 
\item{3.}J.B. Kogut and D.E. Soper, Phys. Rev. {\bf D 1} (1970) 2901.

\item{4.}S.J. Brodsky and H.C. Pauli, {\it Light-cone Quantization and 
QCD}, Lecture Notes in Physics, vol. 396, eds., H. Mitter et. al., 
Springer-Verlag, Berlin, 1991; 
\item{} S.J. Brodsky and G.P. Lepage, 
 in {\it Perturbative Quantum
Chromodynamics}, ed., A.H. Mueller, World Scientific, Singapore, 1989.
\item{5.} K.G. Wilson, Nucl. Phys. B (proc. Suppl.) {\bf 17} (1990). 
R.J. Perry, A. Harindranath, and K.G. Wilson, 
Phys. Rev. Lett. {\bf 65} (1990)  2959. 
\item{6} R.J. Perry, {\it Hamiltonian Light-Front Field Theory and QCD},  
{\sl Hadron Physics 94}, pg. 120, eds., 
 V. Herscovitz et. al., World Scientific, Singapore, 1995 and the 
 references contained therein. 
 
 \item{7.} D. Gross, J. Harvey, E. Martinec and R. Rohm, Nucl. Phys. {\bf 
 B256} (1985) 253. 

\item{8.}H. Lehmann, Nuovo Cimento {\bf 11} (1954) 342. See also 
7.
\item{9.}P.A.M. Dirac, {\it Lectures 
in Quantum Mechanics}, Benjamin, New York, 1964; E.C.G. Sudarshan and 
N. Mukunda, {\it Classical Dynamics: a modern perspective}, Wiley, N.Y., 
1974; A. Hanson, T. Regge and C. Teitelboim, {\it Constrained 
Hamiltonian Systems}, Acc. Naz. dei Lincei, Roma, 1976.
\item{10.} S.S. Schweber, {\it An Introduction to 
Relativistic Quantum Field Theory}, Harper and Row, Inc., New York, 1962. 
\item{11.} {\it Da Teoria Qu\^antica de Campos Sobre a Frente de Onda da 
Luz}, Tese para 
Concurso P\'ublico para Professor Titular do Instituto de F\|sica da 
{\sl Universidade do Estado do Rio de Janeiro}, 
March 1995 {\it } and the earlier references cited here. 
\item{12.}  H.C. Pauli and S.J. Brodsky, Phys. Rev. {\bf D32} (1985) 1993. 
   
\item{13.}  P.P. Srivastava, {\it On spontaneous symmetry breaking 
mechanism in light-front quantized 
field theory}, Ohio State University preprint 91-0481, {\sl SLAC} database no. 
PPF-9148, November 1991. 
\item{14.} P.P. Srivastava, {\it Spontaneous symmetry breaking mechanism 
in light-front quantized field theory} - {\it Discretized formulation},  
 Ohio State University preprint  92-0173, {\sl SLAC} PPF-9222, April 1992, 
 Contributed 
to {\it XXVI Intl. Conference  on High Energy Physics, Dallas, Texas},  
August 92, (AIP Conference Proceedings 272, pg. 2125 Ed. J.R. Sanford ), 
hep-th@xxx.lanl.gov/ 9412193;
\item{} P.P. Srivastava, {\it Light-front Quantization and SSB},  
{\sl Hadron Physics 94}, pg. 253, eds., 
 V. Herscovitz et. al., World Scientific, Singapore, 1995 (
 hep-th@xxx.lanl.gov. nos. 9412204 and 205); 
\item{} P.P. Srivastava, Nuovo Cimento {\bf A107} (1994) 549; 
\item{} see also {\it Lectures } given at {\it XIV ENPC, Caxambu, MG, 1993} 
(hep-th@xxx.lanl.gov/ 9312064). 
\item{15.} P.P. Srivastava, Padova University preprint, DFPF/93/TH/18, 
March 93;  Nuovo Cimento {\bf A108} (1994) 35 ( hep-th@xxx.lanl.gov/9412240). 
\item{16.} B. Simon and R.B. Griffiths, Commun. Math. Phys. {\bf 
33} (1973) 145;  B. Simon, {\it The ${P(\Phi)_{2}}$ Euclidean (Quantum) 
Field Theory}, Priceton University Press, 1974. 
\item{17.} P.P. Srivastava, {\it  Light-front Dynamics of 
Chern-Simons Systems}, ICTP Trieste preprint, IC/305/94 (
 hep-th@xxx.lanl.gov/9412239); see also 
 \item {} invited talk {\it  Light-front Quantization of 
Chern-Simons Systems}
presented at the {\it  International  Workshop on Particle Theory and 
Phenomenology},  
Iowa State University, Ames, May 1995, to be published in the Proceedings, 
World Scientific, Singapore. See also Europhys. Letts. {\bf 33} (1996) 423. 

\item{18} R. Banerjee, A. Chatterjee, and V.V. Sreedhar, Ann. Phys. 
(N.Y.) {\bf 222} (1993) 254; 
\item{} A. Foerster and H.O. Girotti, Phys. Lett. {\bf B 230} (1989) 
83. 
\item {19.} G.W. Semenoff, Phys. Rev. Lett. {\bf 61} (1988) 517.

\item{20.}  C. Hagen, Ann. Phys. (N.Y.) {\bf 157}(1984) 342. 
\item{21} A. Bassetto, O. Nardelli and R. Soldati, {\it YM Theories in 
Algebraic Non-covariant Gauges}, World Scientific, Singapore, 1991. 
\bye